# Tunable Birefringence in Silica Mediated Magnetic Fluid

Urveshkumar Soni, Nidhi Ruparelia, Abhay Padsala, Rucha P Desai[a]

P. D. Patel Institute of Applied Sciences, Charotar University of Science and Technology (CHARUSAT), CHARUSAT Campus, Changa, 388421, Gujarat, India.

[a] Authors to whom correspondence should be addressed: ruchadesai.neno@charusat.ac.in

**ABSTRACT**

The present study reports magnetic and optical properties of silica mediated lauric acid stabilized magnetic fluids. The tunable birefringence ($\Delta n$) and other properties are investigated as a function of (i) concentrations of silica suspension, and (ii) saturation magnetization ($M_S$) 0.5099 kA/m (FN30) and 1.2855 kA/m (F30) of magnetite magnetic fluid (MF). The study reveals that $\Delta n$ suppresses on addition of silica in FN30, whereas enhances (up to critical concentrations of silica) in F30. The magnetic field induced chain observed in the FN30 based fluids are long, thick and scattered, while short, thin and dense chains emerges in F30 based fluid. The magnetic field induced assembly and the magnetic parameters correlates with the results of $\Delta n$. The particle size analysis indicates increment of particle size on addition of silica nanoparticles. The thermogravimetry analysis confirms the direct interaction of silica nanoparticles and the lauric acid coated magnetite particles. This is the first report of direct interaction of silica - magnetite magnetic fluids, and its subsequent effect on tunable birefringence and other properties.

## 1. INTRODUCTION

A stable colloidal suspension of superparamagnetic nanoparticles sterically stabilized with surfactant (either single layer, double layer or multilayer) and dispersed in magnetically passive medium (e.g., water, buffer, kerosene, oil, etc.) forms magnetic fluid (MF) (or ferrofluid)[1][2]. This opaque dark liquid forms magnetic field induced self-assembly. The assembled structure is responsible for many magnetic field induced optical properties like birefringence, transmission, magneto-chromatics, refractive index, etc. It has become base for several MF based potential optical devices such as MF gratings[3], switch[4], modulator[5], capacitor[6], limiters[7], sensors[5], and many more.

The magnetic field induced self-assembly, key factor for material to be birefringent, can be tuned by means of changing the composition of superparamagnetic particles, by varying magnetic volume fraction, by addition of non-magnetic nano/micron-sized particles, etc. It is known that the addition of microparticles of nonmagnetic materials (e.g. silica, carbonyl iron, etc. ) in the magnetic fluid enhances the magnetic field induced structure formation[8,9]. This type of fluid also resembles to magnetorheological (MR) fluids. However, ambiguity in the addition of nonmagnetic nanoparticles have been reported extensively. For example, addition of nano-silica (~10 nm) in water-based MF suppresses the birefringence[10,11], while addition of latex particles (42 nm to 210 nm) in MF enhances chain/column formation[12]. Earlier, we reported the augmentation of chain formation on the addition of halloysite nanotubes (HNTs) [13], and increase in magneto-viscous properties with silica nanoparticles[14]. Moreover, the interaction between the magnetic fluid and non-magnetic objects (HNTs or silica) was not established. Hence, still it is ambiguous that why addition of nonmagnetic particles alter the properties and what type of interaction, if exist, governs the fluidic properties? Here, we report the evidence of direct interaction between the silica nanoparticles and lauric acid coated magnetite particles, and its subsequent effect of the magnetic and optical properties of magnetic fluids. The fluid remains stable on the addition of silica nanoparticles (NPs), which is a plus point for the potentiality of development in any application.

## 2. EXPERIMENTAL
## 2.1 SAMPLE PREPARATION

A magnetite magnetic fluid sterically stabilized with double layer of lauric acid (LA) has been synthesized using chemical co-precipitation route[15]. A single-phase spinel ferrite FCC structure of magnetite, without the presence of any secondary phase has been determined based on x-ray diffraction pattern analysis. It has crystallite size of $(8.2 \pm 0.2)$ nm. This synthesized fluid has been diluted 25 times for the optical measurements and coded here as F30. The magnetic volume fraction ($\varphi_m$) of F30 is 0.0026 (i.e. 0.26%). Further dilution of F30 using lauric acid stabilized solution (0.8 % LA and 5% ammoniated solution (25%) in distilled water) was ultrasonicated for 5 minutes at $45°C$. This results into stable diluted magnetic fluid coded as FN30. The dilutions are stable for more than six months.

We used a colloidal silica AM-30 (LUDOX) (Make: Sigma-Aldrich), consists of silica and aluminum, suspension stabilized using sodium counter ions with the average silica particle size 12nm.

In the F30 and FN30 magnetic fluids fixed amount of AM-30 silica suspension has been added by keeping the total volume constant. Various samples are prepared as follows: F30 MF (99 µl) and 1 µl silica suspension lead to form FA1 fluid. And FN30 MF (99 µl) and 1 µl silica suspension lead to form FNA1 fluid. Hence, the samples are coded as FAX and FNAX depicting the use of F30 and FN30 MF. Similar way other samples were prepared. The list of samples are as follows: FA0.5, FA1, FA1.5, FA2, FA2.5, FA3.5, and FNA1, FNA2, FNA3.

Samples separately prepared for the particles size analysis and thermogravimetric analysis, and coded as FNA0.1 and FA0.27 respectively (described below).

## 3. CHARACTERIZATION
### 3.1 PARTICLE SIZE ANALYSIS

Sample preparation, an essential part of colloidal nanoparticle size analysis, has been carried out as follows. Initially, 1 ml surfactant solution containing 0.1 % LA in 5% ammonia solution (25%) was diluted by adding 1 ml distilled water. In this solution, 0.5 µl FN30 fluid was added. To observe the effect of silica NPs 0.1µl AM-30 suspension was added in the later system. The measurement was carried out using the particle size analyzer (Make: Malvern, Model: Zetasizer, S90) at 25 $^{O}$C. The measurements were repeated for five times to have better statistical average, where each run contains 25 scans.

### 3.2 THERMOGRAVIMETRY

Temperature dependent surfactant decomposition and phase transitions have been determined using thermogravimetric analyzer (Make: Mettler Toledo, Model: TGA/DSC-1) for magnetic fluid (i) without silica (F30) and (ii) with silica suspension (FA0.27). Similar to the particle size analysis measurement, here nominal silica concentration has been used in order to derive the effect of interaction. The samples were dried in a hot air oven at 100 $^{O}$C overnight. The dried powder (~11 mg) sample was taken in an alumina crucible for the measurement. Under the $N_2$ atmosphere, the mass loss was recorded in the temperature range of 50-1000 $^{O}$C in the following segments. Here, the first data indicates temperature range of measurement, and the second shows temperature interval: (i) 50-200 $^{O}$C; 10 $^{O}$C /min., (ii) 200-360 $^{O}$C; 5 $^{O}$C /min., (iii) 360-690 $^{O}$C; 10 $^{O}$C /min. (iv) 690-750 $^{O}$C; 5 $^{O}$C /min., (v) 750-820 $^{O}$C; 2 $^{O}$C /min., and (vi) 820-1000 $^{O}$C; 5 $^{O}$C /min. It took ~ 163 minutes to complete each measurement. The first-order derivative was calculated based on the mass loss data, and was used to determine surfactant decomposition and phase-transition temperature.

### 3.3 MAGNETIZATION MEASUREMENTS

Vibrating sample magnetometer (Make: Lakeshore, Model: 7404) used to perform magnetic measurements at 300K. The following magnetic field range followed by the field interval was set: (a) 0 to 100 G; 1G, (b) 100 to 1000 G, 10G, (c) 1000 to 2500 G, 50G, (d) 2500 to 10000 G; 200G, and (e) 1000 to 12000 G; 50G. The data has been used to determine the initial susceptibility, saturation magnetization, mean magnetic size and size distribution, etc.

### 3.4 MAGNETIC FIELD INDUCED BIREFRINGENCE ($\Delta n$)

The setup comprises of linearly arranged diode laser (unpolarized – 5 mW power), iris diaphragm, polarizer, MF, electromagnet, analyzer, and photo detector (Make: Thorlab Model: DET10A/M). A laser beam passes through an iris diaphragm placed at a distance 0.035 m to remove extra scattering. This focused beam was polarized using a polarizer kept at a distance 0.064 m from iris diaphragm. The polarization angle was +45° with respect to the applied magnetic field (H). The distance between the polarizer and magnetic fluid sample was 0.195 m, whereas, the distance from the MF to analyzer was 0.24 m. The resultant laser beam reached to the photodetector placed at a distance 0.10 m from the analyzer. Between the two pole pieces of electromagnet, a sample cell was placed. The magnetic flux lines were perpendicular to the MF and laser beam. A constant current power supply controlled the magnetic field. The E⊥H configuration was attained by crossing the analyzer and polarizer. The field dependent changes were recorded by adjusting the analyzer angle to maximum intensity ($I_{max}$) and minimum intensity ($I_{min}$). The value of magnetic field induced birefringence ($\Delta n$) determined as

$$\Delta n = \sin^{-1}\left(\frac{2\sqrt{I_{min}/I_{max}}}{1+(I_{min}/I_{max})}\ ch(h_1 - h_2)\right)\frac{\lambda}{2\pi d}$$

The thickness of cell (d) and wavelength of diode laser ($\lambda$) were 120 μm and 650 nm respectively. Here, $h_i$ (i=1,2) is respectively the electric field absorption coefficient for polarized light ∥$^{el}$ and ⊥$^{er}$ with respect to the magnetic field (H). The $h_i$ (i=1, 2) was obtained by solving $I_i = I_{oi}\ e^{-2h_i(H)}$ with the transmitted intensities of the sample $I_{oi}$ at $H = 0\ T$, and $I_i$ at $H \neq 0\ T$. Thus, magnetic field dependent $\Delta n$ was determined based on $I_{max}$, $I_{min}$, $I_{oi}$, and $I_i$.

### 3.5 OPTICAL MICROSCOPY

An inverted microscope (Make: Meiji Techno Model: IM7100) with a 20 × objective lens and numerical aperture (NA) 0.4 attached to charged coupled device (CCD) camera (Make: Jenoptik), operated using ProgRes-C3 software, has been used to record magnetic-field-induced structure formations. The image area was pre-calibrated using the standards provided by the manufacturer. The images were captured in E⊥H configuration. The sample was prepared by sandwiching the magnetic fluid between a glass slide and coverslip. Rare earth cylindrical magnet was used to provide a constant magnetic field (H) ~$0.055\ T$.

# 4. RESULT AND DISCUSSION

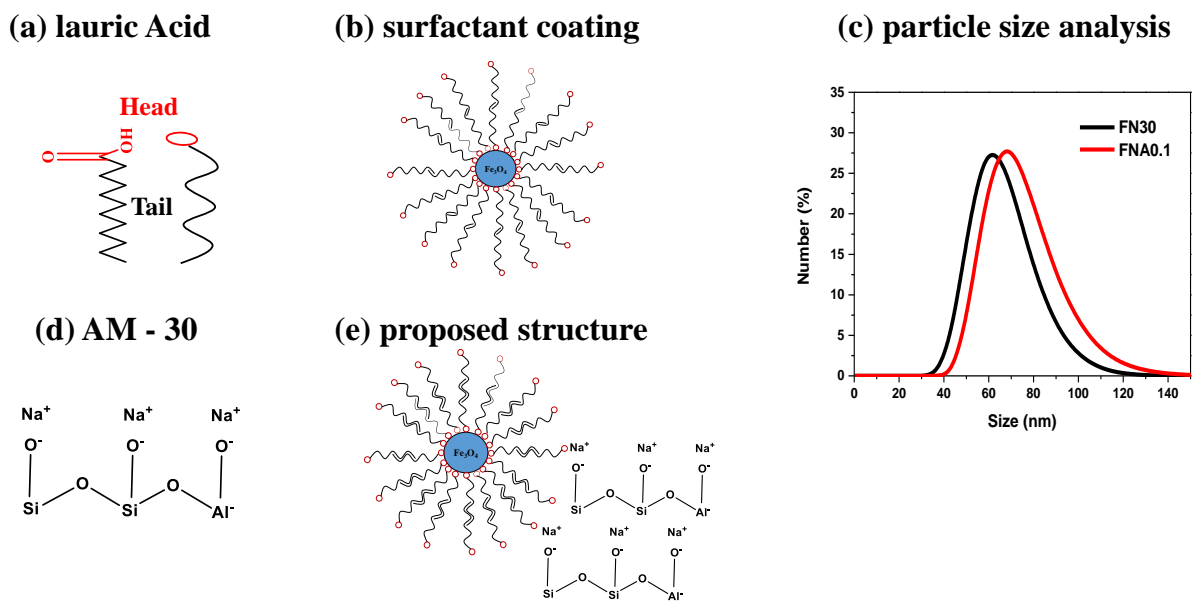

Figure 1: Chemical structures of (a) lauric acid (b) magnetite particle with primary and secondary coating of LA, (c) results of particle size analyzer, chemical structure of (d) silica suspension AM-30 and (e) interaction of silica NPs with LA coated magnetite.

Figure 1(a) shows the chemical structures of LA containing a hydrophilic head and a hydrophobic tail. During the reaction, at moderately high temperature (80-90 $^{O}$C), COOH group reduces to COO$^{-}$ and attaches to the positively charged nanoparticle, leaving an unbound hydrophobic tail. In the polar medium second layer of LA provides stability towards agglomeration (figure 1(b)) in reverse configuration (tail to head). Figure 1(c) shows hydrodynamic particle size distribution (number distribution) for magnetic fluid without and with silica nanoparticles, as 58 ($\pm$1) nm and 68 ($\pm$1) nm respectively. Assuming the mean magnetite particle size 12 nm and lauric acid chain length 2 nm, the expected total hydrodynamic size is around 16 nm. Noted here that the average hydrodynamic size is high. For the particle size measurement, the magnetic fluid has been diluted many folds with the additional free surfactant compared to the fluids used for the measurements (F30 & FN30). During the extensive dilution, one may not overlook the possibility of formation of dimer/trimers, which eventually decorated by free lauric acid that forms multilayer around the particle(s). The interest of this measurement was not to determine exact particle size, but to understand the interaction of the silica with the lauric acid stabilized magnetite nanoparticles. In case of non-interacting behavior of silica, one should expect two peaks (i.e. ~ 12nm of silica and 58 nm of coated magnetite). On the contrary, single peak around 68 nm is observed. As per the datasheet the size of silica NPs is 12 nm, whereas observed enhancement is ~ 10 nm. The difference of 2 nm could be due to the compression of surfactant in the presence of silica NPs. It is quite possible that silica NPs either interacting with free lauric acid molecules or lauric acid-coated magnetite particles. Further, to recognize the nature of interaction, thermogravimetric measurement has been carried out.

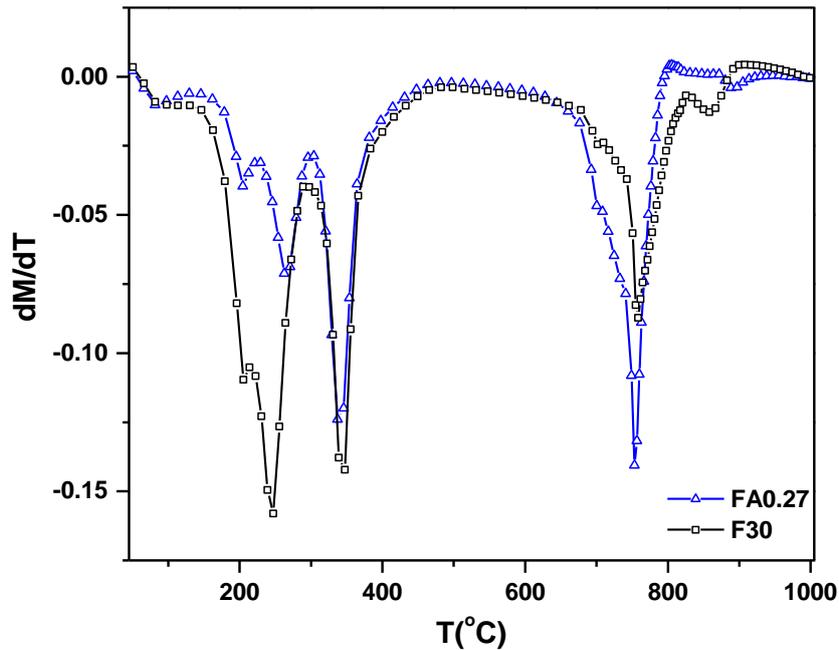

Figure 2: First order derivate of temperature dependent mass loss for FA0.27 & LA100 fluids.

Figure 2 shows typical data of the first derivative of mass loss $\left(\frac{dM}{dT}\right)$ for F30 and FA0.27 fluids, with major six peaks. The mass loss observed up to ~120 $^{O}$C in both the samples is attributed to absorbed water. The second peak at 200 ± 5 $^{O}$C in both the systems indicates decomposition of free surfactant. The difference observed in the mass loss, i.e. ~ 11% in F30 and ~ 4% in FA0.27 is assigned to the interaction of silica NPs with free surfactant. The next peak observed at 247 ± 5 $^{O}$C (~ 15% mass loss) in F30 ascribes decomposition of the physi-adsorbed layer of LA on magnetite particles. This peak shifts to 262 ± 5 $^{O}$C (~ 7% mass loss) in FA0.27 suggesting interaction of silica nanoparticles with the secondary layer (physi-adsorbed) of LA. In both the peaks, the reduction of mass loss in presence of silica nanoparticles attributed to the redistribution of surfactant around the surface of silica NPs. The decomposition of chemi-adsorbed LA occurs at 347 $^{O}$C in both the fluids. The constant peak position indicates the interaction of silica NPs with the secondary layer of surfactant. The temperature >600 °C shows the phase transition of magnetite. The peak observed in F30 at ~ 758 ± 2 °C is attributed to the phase transition of FCC spinel ferrite $Fe_3O_4$ to wustite Fe-O[16]. Further, ~ 858 ± 5 °C is attributed to the transformation to metallic Fe. In the silica NPs added fluid, similar phase transition with little difference is observed. Before the decomposition of $Fe_3O_4$ structure, the peak becomes broaden followed by evident transformation at 753 ± 2 °C. It is an interesting observation, it may be due to the interaction of silica NPs with the bare $Fe_3O_4$ (surfactant decomposed <600 °C) and forms a new composite. The phase transition observed at 758 ± 2 °C is ascribed to the decomposition of thermally formed silica-Fe phase. In summary, TGA result confirms the interaction of silica NPs with the secondary layer of surfactant around $Fe_3O_4$ particles. The results also support the inference of particle size analysis. Figure 1(d & e) shows the chemical structure of AM-30 silica NPs and possible interaction of silica NPs with surfactant coated magnetite particle.

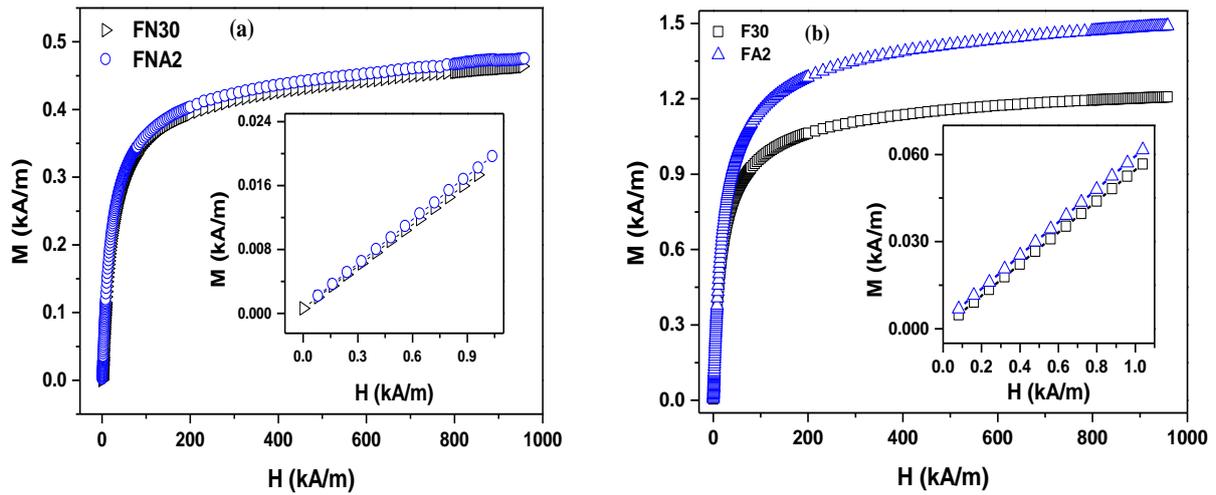

Figure 3: Magnetization measurement of (a) FN30 & FNA2, and (b) F30 & FA2 fluids, with the respective initial susceptibility data (inset).

Figure 3 shows magnetic field ($H$) dependent response of (a) FN30 & FNA2, and (b) F30 & FA2 fluids. Inset in the respective figures represents linear treads at low fields. The initial susceptibility ($\chi_i$) determined by fitting with the linear equation is listed in the Table-1. The data recorded during the field sweep indicates a linear increase in magnetization in the low field, exponentially increases in the mid-field region - due to spontaneous magnetization - followed by going towards saturation at high field. Table-1 shows the saturation magnetization ($M_s$) determined by extrapolating $M\ vs\ 1/H$ at $1/H = 0$ (i.e., intercept). The table also shows the magnetic volume fraction determined using $\varphi_m = M_s/M_d$, assuming domain magnetization ($M_d$) of $Fe_3O_4$ = 485 $kA/m$ (bulk value). The mean particle diameter ($D_m$) derived using the ideal Langevin initial susceptibility equation, $\chi_i = \frac{\mu_0 \pi M_d^2 D_m^3 \varphi_m}{18 k_B T}$, where, $\chi_i$ is the initial susceptibility, $\mu_0$ the permeability of free space, constant $T$ the absolute temperature and $k_B$ the Boltzmann constant. For the magnetic fluid exhibiting superparamagnetism, Chantrell *et al.* [17] proposed the magnetic particle size and its distribution deduction in the form of volume and number weighted average expressed as,

$$D_{mV} = \left[\frac{18 k_B T}{\mu_0 \pi M_d} \sqrt{\frac{\chi_i}{3 M_s H_0}}\right]^{1/3}, D_{mN} = \left[2 k_B T \sqrt{3 M_s / \chi_i H_0^3} / \pi M_d\right]^{1/3}, \sigma_D = \frac{1}{3}\sqrt{ln\left[\frac{3\chi_i H_0}{M_s}\right]}$$

where, $D_{mV}$ is the mean particle diameter in the case of volume weightage average, $D_{mN}$ the mean particle diameter of number weightage average, $\sigma_D$ the log-normal size distribution parameter, and $H_0$ the field at which the high field data for $M/M_s$ vs $1/H$ extrapolates to $M = 0$. Noted here that the volume average has less uncertainty due to $(1/H_0)$ rather than $(1/H_0)^3$ effect.

Table 1: An initial susceptibility $(\chi_i)$, and saturation magnetization ($M_s$) derived from data. The magnetic volume fraction ($\varphi_m$) calculated assuming constant $M_d$. The mean particle diameter $D_m$ determined based on the initial susceptibility equation. The volume-weighted mean magnetic diameter ($D_{mV}$), number weighted mean diameter ($D_{mN}$) & log-normal size distribution ($\sigma_D$) calculated from equations.

| Sample | $\chi_i$ (± 0.0001) | $M_s$ ($kA/m$) (± 0.0045) | $\varphi_m$ | $D_m$ ($nm$) (± 0.1) | $D_{mV}$ ($nm$) (± 0.01) | $D_{mN}$ ($nm$) (± 0.01) | $\sigma_D$ |
|---|---|---|---|---|---|---|---|
| FN30 | 0.0174 | 0.5099 | 0.0011 | 11.0 | 7.36 | 3.56 | 0.49 |
| FNA2 | 0.0180 | 0.5498 | 0.0011 | 10.8 | 6.74 | 2.81 | 0.54 |
| F30 | 0.0543 | 1.2825 | 0.0026 | 11.8 | 8.23 | 4.31 | 0.46 |
| FA2 | 0.0570 | 1.5965 | 0.0033 | 11.2 | 7.95 | 4.34 | 0.45 |

Table 1 shows magnetic parameters determined for FN30, FNA2, F30, and FA2 fluids. The FN30 was prepared by diluting F30 fluid, the observed difference in $\chi_i$ and $M_s$ indicates the breaking of aggregates on dilution. The $\chi_i$ and $M_s$ increases in FNA2 and FA2 fluids compared to their respective parent fluids. However, the $D_m$ reduces on addition of silica. The $\sigma_D$ increases in FNA2 fluid, which suggests re-distribution of particle due to the interaction with silica NPs. The $\sigma_D$ nominally decreases in FA2 fluid. The effect of diamagnetic silica NPs is apparent in the magnetically diluted system. The magnetic field induced structure formation is expected to be different in all these systems. Hence, the magnetic field induced birefringence and optical microscopy have been carried out.

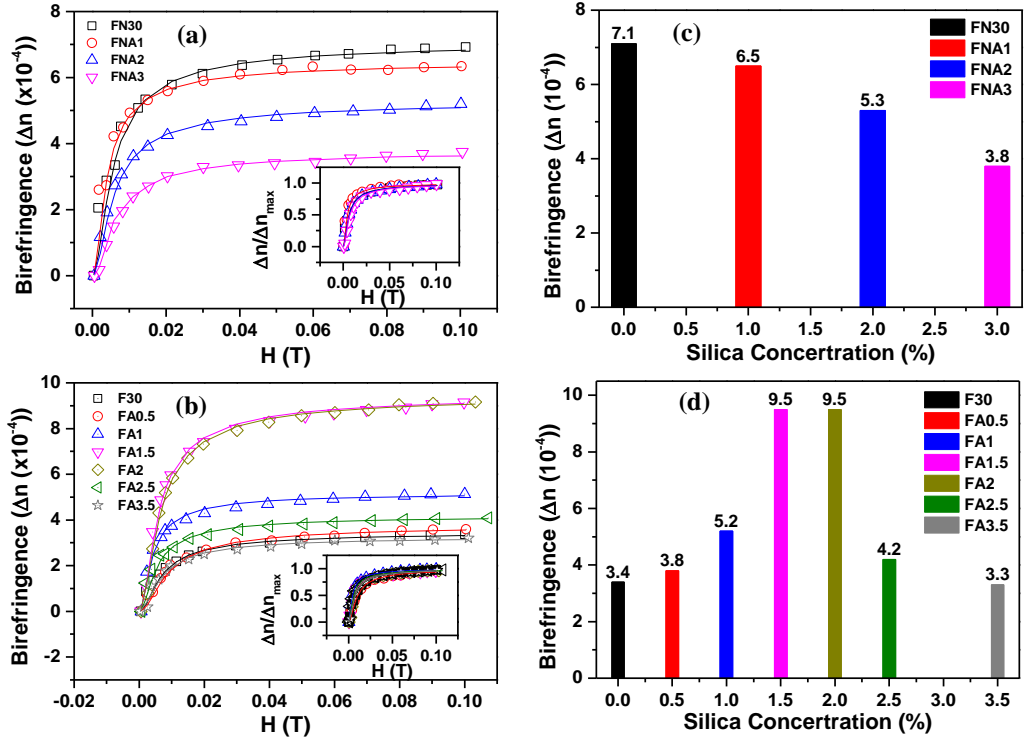

Figure 4: Magnetic field-induced birefringence and reduced birefringence (inset) for FA & FNA systems as a function of magnetic field (a) & (b) and as a function of silica concentration (c) & (d) respectively.

Fig. 4 (a & b) shows magnetic field induced birefringence ($\Delta n$) with respective inset figures of reduced birefringence $\left(\frac{\Delta n}{\Delta n_{max}}\right)$ as a function of the magnetic field for FN30, FNA1, FNA2, FNA3 and F30, FA0.5, FA1, FA1.5, FA2, FA2.5, FA3.5 fluids. It infers from the figure that, with increasing magnetic field, $\Delta n$ spontaneously increases and goes toward saturation. Owing to superparamagnetic nature, the magnetic field dependent birefringence can be explained by incorporating Langevin function at the relatively low magnetic field,

$$\Delta n = \Delta n_{max} \left(1 - \frac{3L(\alpha)}{\alpha}\right)$$

where, $L(\alpha) = Coth(\alpha) - \frac{1}{\alpha}$ is a Langevin function, and $\alpha = \frac{\mu H}{kT}$ the Langevin parameter, with optimized $\Delta n_{max}$. In figure 4 (a & b) open symbol depicts experimental data of fluid while solid line fits above equation. Inset of Figure 4(a & b) normalized birefringence ($\Delta n/\Delta n_{max}$) curves superimpose with each other. Per se it indicates result of the product $\mu H$, which eventually remains unchanged on the addition of silica suspension. This analysis can be refined by incorporating moment distribution function. But that is not the scope of the present work. Figure 4 (c & d) shows variation in $\Delta n_{max}$ with silica NPs concentration for FN30 and F30 fluid parent systems, respectively. It is observed from Figure 4(c) that $\Delta n_{max}$ suppresses on addition of silica NPs. On the contrary, $\Delta n_{max}$ increases with increasing silica NPs concentration reach to maxima and then decreases in F30 based fluids (Figure 4(d)). The suppression behavior agrees with earlier reported results [10,18]. However, the increase in $\Delta n_{max}$ on addition of diamagnetic silica NPs has been reported for the first time here. The birefringence is related to the field induced structure formations, e.g. chain, column, etc.

Figure 5 shows the magnetic field (H = 0.055T) induced microscopic images of FNA & FA systems (typical pictures shown here for the brevity). The images were captured after 1 minute of applying the magnetic field. It is observed that under the influence of the external magnetic field, the particles starts aligning and forms chains in the field direction.

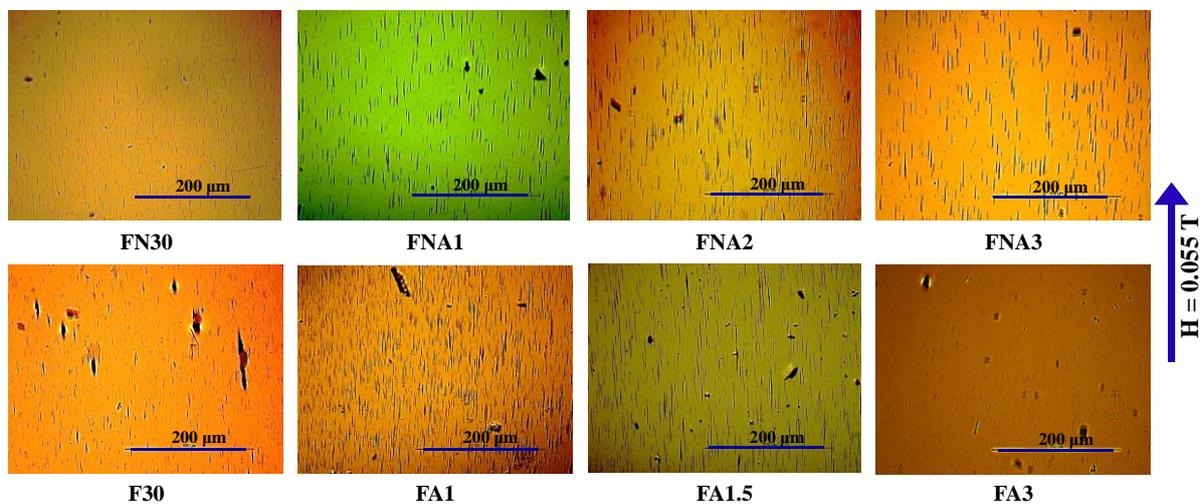

Figure 5: Microscopic confirmation of chain formation on the exposure of 0.055T magnetic field after 1 minute.

Table 2: The analyzed chain parameters, i.e., chain length & chain width along with its respective % count for the typical FNA & FA systems.

|  | FN30 | FNA1 | FNA2 | FNA3 | F30 | FA1 | FA1.5 | FA3 |
|---|---|---|---|---|---|---|---|---|
| **Length-1 (μm)** | 3.7 | 3.7 | 3.7 | 3.7 | 2.9 | 3.7 | 3.7 | 2.9 |
| **% Counts (Count)** | 66% (2693) | 58% (747) | 47% (777) | 43% (711) | 64% (1667) | 41% (961) | 62% (890) | 68% (2045) |
| **Length-2 (μm)** | 7.2 | 10.8 | 11.8 | 15 | 8.8 | 11.9 | 12.3 | 10 |
| **% Counts (Count)** | 10% (280) | 7% (54) | 14% (109) | 21% (152) | 10% (175) | 18% (176) | 12% (108) | 13% (275) |
| **Length-3 (μm)** | 10.6 | 17.7 | 19.8 | 26.1 | 14.8 | 20.1 | 20.9 | 17 |
| **% Counts (Count)** | 6% (155) | 7% (50) | 11% (85) | 10% (74) | 6% (96) | 16% (150) | 7% (63) | 7% (133) |

|  | FN30 | FNA1 | FNA2 | FNA3 | F30 | FA1 | FA1.5 | FA3 |
|---|---|---|---|---|---|---|---|---|
| **Width-1 (μm)** | 3.7 | 3.7 | 3.7 | 3.7 | 2.9 | 3.7 | 3.7 | 2.9 |
| **% Counts (Count)** | 70% (1693) | 56% (747) | 63% (777) | 44% (711) | 56% (1667) | 55% (961) | 68% (890) | 69% (2045) |
| **Width-2 (μm)** | 6.1 | 5.2 | 5.1 | 5.5 | 3.7 | 5.7 | 5.1 | 4.9 |
| **% Counts (Count)** | 14% (372) | 11% (84) | 12% (96) | 22% (157) | 18% (269) | 15% (148) | 11% (295) | 14% (286) |
| **Width-3 (μm)** | 8.5 | 6.6 | 6.4 | 7.2 | 4.6 | 7.6 | 6.5 | 6.9 |
| **% Counts (Count)** | 6% (163) | 12% (86) | 9% (70) | 13% (91) | 8% (139) | 8% (75) | 7% (60) | 5% (109) |

The images shown in the figures have been analyzed with ImageJ software. Table 2 shows the average chain parameters, i.e. length and width, contributing significantly. In FN30, FNA1, FNA2, and FNA3 fluids are having similar chain length, i.e. 3.7 μm, with a systematic decrement in the % counts. The next significant chain length increases with increase in silica concentration. It spans from 7.2 to 15 μm with moderately high % counts. The next large chains observed ranging from 10.6 to 26.1 μm. The chain width ranges from 3.7 to 8.5 μm. Correlating Table 2 and the images of Figure 5, the chains observed in FN30 and FNA fluids are long, thick, and scattered. The interchain distance increases with increasing silica.

The chain length observed in F30, FA1, FA1.5, and FA3 fluids are 2.9, 3.7, 3.7 and 2.9 μm respectively, with nearly similar % counts (except FA1). The next significant chains observed are 8.8, 11.9, 12.3 and 10 μm long with nearly similar % counts (except FA1). The next significant chain length contributing in the process are 14.8, 20.1, 20.9, and 17 μm with above mentioned trend in % counts. The width of the chain ranging from 2.9 to 7.4 μm. It is inferred from chain parameters, that inclusion of silica makes a difference in the chain parameters. Chain length increases up to FA1.5 and then decreases. Correlating chain parameters and the images, the chain observed are short, thin, and dense in F30 based fluids compared to FN30 based fluids. The analyzed chain parameters supports the birefringence analysis and discussed below in detail.

The suppression and the enhancement in birefringence on the addition of silica NPs in the magnetic fluid having two magnetic volume fractions is analyzed as follows. The silica NPs interacts with the magnetite particles, hence system mimics as magnetorheological (MR) fluid. In zero magnetic field, particles exhibit zero net magnetic moment, and the particles are dominated by Brownian motion. The magnetic field dependent magnetic moment of a single particle is given by,

$$m = \frac{\pi}{6} a^3 \chi_{eff} H$$

where, $a$ is the magnetic nanoparticle diameter, and $\chi_{eff}$ the effective susceptibility. The anisotropic dipolar potential energy of pairs of particles is expressed as, $U_{ij}(r_{ij}, \theta_{ij}) = \frac{m^2 \mu_0}{4\pi} \left( \frac{1 - 3\cos^2\theta_{ij}}{r_{ij}^3} \right)$ with $r_{ij}$ center to center distance between the i$^{th}$ and j$^{th}$ particles, and $\theta_{ij}$ the angle between the vector $r_{ij}$ and the magnetic field applied. Using this, the relative strength of dipolar interaction in terms of thermal energy is expressed by coupling parameter $\lambda = -\frac{U(a,0)}{k_B T} = \frac{\pi \mu_0 a^3 \chi^2 H_0^2}{72 k_B T}$. The magnetic particles assemble into the aligned structure ($\lambda \gg 1$) on application of the field, forming dipolar chains and exhibits strong Landu-Peierls fluctuation. Halsey and Toor (HT) illustrates the long-range coupling among dipolar chains due to chain formations, and possess an attractive interaction through power-law decay. The HT model was modified by Martin *et al.*[19],

$$U \sim \frac{m}{a} \langle H^2 \rangle^{1/2} \sim \frac{\chi H (\mu_0 K_B T)^{1/2} a^{1/5}}{\rho^2}$$

Here, $\rho$ is the distance among two chains or columns. This energy can be either attractive or repulsive. The interaction energy per unit length increases with increasing the field and/or decrease in the inter-chain distance resulting in lateral coalesce of two chains. Consequently, the separation distance ($\rho$) increases and lowering U, and resulting into reduction of overall energy of the system.

It is evident from Table 2 that chains observed in FN30 based fluids are long, thick, and scattered, with increasing inter-chain distance. Also, heterogeneous distribution in the chain length and width is observed. Referring to Table-1, the $\sigma_D$ increases from 0.49 to 0.54 on addition of silica, lead to form long and thick chains with increasing inter-chain distance. Correlating the chain parameters with the % counts, $\Delta n_{max}$ suppressing with increasing silica concentrations. The chains observed in F30 based fluids are short, thin, and dense. The chain parameters increase up to critical concentration of FA1.5 followed by trailing behavior. Similar behavior is also observed in the birefringence. Hence, the suppression and enhancement in birefringence are attributed to the magnetic field induced self-assembly, and also combination of magnetic to silica volume fraction.

## 5. CONCLUSION

Generally, addition of micron-sized non-magnetic particles (e.g., silica, latex, etc.) in the magnetic fluid enhances the magnetooptical properties resembling magnetorheological (MR) fluids[20]. On the other side, ambiguity in the magnetooptical properties of non-magnetic nanoparticles added magnetic fluids have been observed. For example, [10,11] demonstrates that addition of silica nanoparticles suppresses the magnetic field induced birefringence and structure formations, while, [12] reports enhancement. However, ambiguity in the magnetic field induced change in the properties are not discussed yet. Also, the nature of interaction with the magnetic particles is not discussed much. The present study focuses on the magnetic and optical properties of colloidal silica (~12nm) nanoparticles added lauric acid stabilized magnetic fluids. The study carried out by varying (i) concentrations of silica suspension, and (ii) saturation magnetization ($M_S$) 0.5099 kA/m (FN30) and 1.2855 kA/m (F30) of magnetite magnetic fluid (MF). The crystallite size of magnetite is 8.4 nm. Double layer of lauric acid surfactant will increase the size of the particles, but remains in nano-regime. The hydrodynamic particle size determined for much diluted magnetic fluid is ~ 58 nm. Addition of silica in a typical concentration increase this size to ~ 68 nm. This is the first evidence of interaction of silica with lauric acid stabilized magnetite magnetic fluid. Further, thermogravimetric study reveals the shift in the physi-adsorbed layer of lauric acid in presence of silica nanoparticles. This shift confirms that silica nanoparticles interact with the second layer of lauric acid, which provides stability to magnetite in the aqueous media. The magnetization measurements indicates increment in the saturation magnetization, however, nominal decrease in the mean magnetic size of FN30 based fluid, and nominal increase in the F30 based fluid were observed in presence of silica nanoparticles. Also, the log-normal size distribution function $\sigma_D$ increases in FN30 based fluid, whereas it decreases in F30 based fluid. Addition of silica suspension suppresses magnetic field induced birefringence (Δn) in FN30 based fluid. Interestingly, Δn increases up to critical concentrations of silica suspension in F30 based fluid, and then it decreases. The microscopic observations of magnetic field induced assembly indicates that the chains formed in FN30 based fluids are long, thick and scattered, while in F30 based fluids chains formed were comparatively short, thin, and dense. The orientation of chains and the assembled structure does influence on the birefringence of magnetic fluid. It is summarized that both, the suppression and enhancement, does depend on the ratio of silica and magnetite particles volume fraction. The tunability of these properties are dedicated to the interaction of silica with the lauric acid coated magnetite nanoparticles. This is first report establishing the direct interaction with the silica and magnetite nanoparticles via lauric acid surfactant, and its influence on the magnetic field induced properties.

**ACKNOWLEDGMENT**

This work was carried out under a grant No: EMR/2016/002278 sponsored by Science and Engineering Research Board (SERB), Department of Science and Technology (DST), India. The initial measurements of birefringence have been carried out by Abhay Padsala as a part of M.Sc. (Physics) dissertation work. We acknowledge guidance provided by Prof. Arbindo Ray for TGA analysis. We thanks to Dr. Kinnari Parekh for the help in magnetization measurement.